\documentclass{iopart}
\usepackage{graphics}
\usepackage{cite}

\newcommand{\ket}[1]{\left\vert #1\right\rangle}
\newcommand{\bra}[1]{\left\langle #1\right\vert}
\newcommand{\ketbra}[2]{\left\vert #1 \middle\rangle\middle\langle #2\right\vert}
\newcommand{\braket}[2]{\left\langle #1 \middle\vert #2\right\rangle}
\newcommand{\sech}{\mathrm{sech}}
\begin{document}

\title[Tripartite Coherent-Entangled State, Generation \& Applications]{A New Tripartite Coherent-Entangled State Generated by An Asymmetric Beam-Splitter and Its Applications}
\author{ Ma Xu$^\mathrm{a,b}$, Zhang Shuang-Xi$^\mathrm{a}$, Dong
Chuan-Fei$^\mathrm{a,c}$
\\
{$^\mathrm{a}$\small Department of Modern Physics, University of
Science and Technology of China (USTC), Hefei, Anhui 230026,
P.R.China}
\\
{$^\mathrm{b}$\small Department of Physics, Syracuse University,
Syracuse, NY 13244, U.S.A.}
\\
{$^\mathrm{c}$\small School of Earth and Atmospheric Sciences,
Georgia Institute of Technology, Atlanta, Georgia 30332, U.S.A.} }

\ead{shuangxi@mail.ustc.edu.cn}

\begin{abstract}
A new kind of tripartite coherent-entangled state (CES)
$\ket{\beta,\gamma, x}_{\mu\nu\tau}$ is proposed, which exhibits the
properties of both coherence and entanglement. We investigate its
completeness and orthogonality, and find it can make up a
representation of tripartite CES. A protocol for generating the
tripartite CES is proposed using asymmetric beam splitter.
Applications of the tripartite CES in quantum optics are also
presented.
\end{abstract}

\maketitle

\section{Introduction}
\label{introduction}
\par
The concept of entanglement, originated by Einstein, Podolsky and
Rosen (EPR) for arguing the incompleteness of quantum mechanics,
plays a key role in understanding some fundamental problems in
quantum mechanics. Quantum entangled states have been vastly studied
by physicists due to their potential usage in quantum information
and quantum communication. In a quantum entangled system, a
measurement performed on one part of the system provides information
about the remaining part, and this is now known as the basic feature
of quantum mechanics, weird though it seems. For a good
understanding entanglement, Ref. \cite{1} will be useful. Beyond all
entangled states, the continuous variable entangled states are of
great application in quantum optics and atomics area, where the
continuous variables are just the quadrature phase of optics field.
Detail acquaintance of continuous variable can refer to Ref.
\cite{2,3}. And it can be inferred that continuous variable
entanglement states such as topological entanglement may play an
important roll in understanding famous and obscure phenomenons in
low temperature physics such as the fractional electron charge
effect \cite{4}. On the other hand, the theoretical research has go
ahead of experiment to construct various continuous variable
entangled states: idealized EPR state $\ket{\eta}$\cite{5}, two mode
coherent entangled state $\ket{\alpha,x}$ \cite{6}, and arbitrary
multi-mode entangled state\cite{7} and so on. All these mentioned
states have been constructed and property-analyzed basing on IWOP
technique\cite{8,9,10,11}. Contrast to classical quantum optical
states, these states present nonclassical properties such as partial
non-positive of Wigner distributive and Mandel factor\cite{12},
divergence in special point in phase space of Glauber-Sudarshan
representation \cite{13}. Among these states, CES is of special
interesting due to its intrinsic nature of the merging of coherence
and entanglement. As far as we know, multipartite CES have been
write down in Ref. \cite{6,15}. Here we propose a new tripartite CES
$\ket{\beta,\gamma, x}_{\mu\nu\tau}$, which is the generalization
version of the old tripartite CES $\ket{\beta,\gamma, x}$. This
generalization is not trivial. As tripartite CES, we find that it
can play as the continuous base in Hilbert space of
square-integrated property, after checking its completeness and
orthogonality.

\par
This paper is arranged as follows, in Sec. \ref{triparite} the
explicit form of tripartite CES $\ket{\beta,\gamma, x}_{\mu\nu\tau}$
is present in Fock space by virtue of the technique of integration
within an ordered product of operators (IWOP), and then some main
properties are analyzed in Sec. \ref{properties}. The protocol for
generating the tripartite CES is proposed In Sec.
\ref{generating-CES} using asymmetric beam splitter. Sec.
\ref{application} is devoted to briefly discussing some potential
applications of $\ket{\beta,\gamma, x}_{\mu\nu\tau}$ in quantum
optics. A brief conclusion is presented in the last section.

\section{The Introduction of tripartite CES}
\label{triparite}
\par
For a physical state, it is hoped that it can span a complete space.
For example, the Fock state and the coherent state are both
complete. It has been shown \cite{7} that by constructing
miscellaneous normally ordered Gaussian integration operators, which
are unity operators, and then considering their decomposition of
unity we may derive new quantum mechanical states possessing the
completeness relation and orthogonality. For example, from the
normally ordered Gaussian form of unity
\begin{equation}
\int\frac{\rmd ^2 z}{\pi}:\exp\left[-(z^*-a^\dag)(z-a)\right]:=1
\label{1}\\
\end{equation}
and using the normal ordering form of vacuum state projector
$\ketbra{0}{0}=:\exp(-a^{\dag}a):$\cite{16,17}, where $:\ :$ is
signal of normal ordering, we can make the decomposition
\begin{equation}
:\exp\left[-(z^*-a^\dag)(z-a)\right]:=\ketbra{z}{z} \label{2}
\end{equation}
so the form of coherent state $\ket{z}=\exp(-\frac{\vert
z\vert^2}{2}+z a^\dag)\ket{0}$ emerges. Similarly, by examining
\begin{equation}
\int\frac{\rmd ^2
\eta}{\pi}:\exp\left[-(\eta^*-a_1^\dag+a_2)(\eta-a_1+a_2^\dag)\right]:=1
\label{two}
\end{equation}
and decomposing the integrand in \Eref{two} we observe the emergence
of bi-particle ideal EPR state $\ket{\eta}=\exp[-\frac{1}{2}\vert
\eta \vert^2 + \eta a_1^\dag - \eta^* a_2^\dag + a_1^\dag
a_2^\dag]\ket{00}$ \cite{5}. And if we go further, by decomposition
the following unity of the Gaussian operator integration within
normal ordering
\begin{eqnarray}
\int_{-\infty}^{\infty}\frac{\rmd x}{\sqrt{\pi}} \int\frac{\rmd^2
\alpha}{2\pi}:\exp\left[-\left(x-\frac{\mu X_1+ \nu
X_2}{\sqrt{2}\lambda}\right)^2\right]\nonumber\\
\times\exp\left\{-\frac{1}{2}\left[\alpha^*-\frac{1}{\lambda} (\nu
a_1^\dag -\mu a_2^\dag)\right]\left[\alpha-\frac{1}{\lambda} (\nu
a_1 -\mu a_2)\right]\right\}:=1
\end{eqnarray}
after the decomposition, we can get the expression of the state
\begin{eqnarray}
\ket{\alpha,x}_{\mu\nu}&=&\exp\left[-\frac{1}{2}x^2-\frac{1}{4}\vert
\nu \alpha \vert^2 +\lambda\alpha a_1^\dag +\frac{\mu}{\lambda}
\left(x-\frac{\alpha\mu}{2}\right)a_1^\dag\right.\nonumber\\
&&\phantom{\exp[}\left.
+\frac{\nu}{\lambda}\left(x-\frac{\alpha\mu}{2}\right)a_2^\dag-\frac{1}
{(2\lambda)^2}\left(\mu a_1^\dag + \nu a_2^\dag
\right)^2\right]\ket{00} \label{bipartite-CES}
\end{eqnarray}
This is the new bipartite CES proposed in Ref. \cite{18}, with
$\mu^2+\nu^2=2\lambda^2$. Using the bosonic communicative relation
$[a_i,a_j^\dag]=\delta_{ij}$, we have \numparts
\begin{eqnarray}
a_1 \ket{\alpha,x}_{\mu\nu} &=&\left[
\alpha\lambda+\frac{\mu}{\lambda} \left(
x-\frac{\alpha\mu}{2}\right) -\frac{\mu}{2\lambda^2}(\mu a_1^\dag
+\nu a_2^\dag)\right]\ket{\alpha,x}_{\mu\nu}\\
a_2 \ket{\alpha,x}_{\mu\nu}& =&\left[\phantom{\alpha +\lambda}
\frac{\nu}{\lambda} \left( x-\frac{\alpha\mu}{2}\right)
-\frac{\nu}{2\lambda^2}(\mu a_1^\dag +\nu
a_2^\dag)\right]\ket{\alpha,x}_{\mu\nu} \label{5}
\end{eqnarray}
\endnumparts
which satisfy the following eigenequations
\numparts
\begin{eqnarray}
\frac{1}{2}(\mu X_1 + \nu
X_2)\ket{\alpha,x}_{\mu\nu}&=&\frac{\lambda
x}{\sqrt{2}}\ket{\alpha,x}_{\mu\nu}\\
(\nu a_1 - \mu a_2)\ket{\alpha,x}_{\mu\nu}&=& \nu \alpha \lambda
\ket{\alpha,x}_{\mu\nu},\label{6}
\end{eqnarray}
\endnumparts
which means $\ket{\alpha,x}_{\mu\nu}$ is actually the common
eigenvector of $(\mu X_1 + \nu X_2)$ and $(\nu a_1 - \mu a_2)$, and
$[(\mu X_1 + \nu X_2),(\nu a_1 - \mu a_2)]=0$, and
\begin{equation}
X_i=(a_i+ a_i^\dag)/\sqrt{2},\qquad
P_i=(a_i-a_i^\dag)/(\sqrt{2}\rmi) \label{xp}
\end{equation}
However, if we want to obtain the expression of tripartite CES, it
may become tedious to construct such a complex quadratic gaussian
polynomial of three-mode of generate operator in entangled form to
derive tripartite CES. Fortunately, we can stride over this problem
just oppositely, first constructing the tripartite counterpart
formally analogue to bipartite CES, then checking it satisfies the
similar relationship \Eref{two}. Along this way, tripartite CES
$\ket{\beta,\gamma, x}_{\mu\nu\tau}$ can be introduced as
\begin{eqnarray}
\fl \ket{\beta,\gamma, x}_{\mu\nu\tau}=\exp & \left\{
\left[-\frac{3}{4}x^2 -\frac{1}{6\nu} (\beta^*\gamma+
\beta\gamma^*)\mu\tau^2 -\frac{1}{6}\vert\gamma\vert^2\tau^2
\left(1+\frac{\mu^2}{\nu^2}\right) \right. \right.\nonumber\\
&\quad - \left.\frac{1}{6}\vert\beta\vert^2
\left(\nu^2+\tau^2\right)\right]+ \frac{1}{3\lambda}
\left[\left(\beta
(\nu^2+\tau^2)+\frac{\gamma\mu\tau^2}{\nu}+3x\mu\right)a_1^\dag\right.\nonumber\\
&\quad +
\left(-\beta\mu\nu+\gamma\tau^2+3x\nu\right)a_2^\dag\nonumber\\
&\quad + \left.\left(-\frac{\gamma(\mu^2+\nu^2)\tau}{\nu}-\beta\mu\tau+3x\tau\right) a_3^\dag\right]\nonumber\\
&\quad  \left.-\frac{1}{6\lambda^2}\left( \mu a_1^{\dag}+\nu
a_2^{\dag} +\tau a_3^{\dag}\right)^2 \right\}\ket{000} \label{7}
\end{eqnarray}
where $\mu$, $\nu$, $\tau$ are three independent parameters, and
$3\lambda^2=\mu^2+\nu^2+\tau^2$, this identity hold to make sure it
can be generated by beam-splitter which will be instructed in Sec.
\ref{generating-CES}. In particular, when $\mu=\nu=\tau=1$, \Eref{7}
will reduce to
\begin{eqnarray}
\fl \ket{\beta,\gamma, x}=\exp & \left\{-\frac{3}{4}x^2 -\frac{1}{6}
\left(\beta\gamma^* +\beta^*\gamma+ 2\vert\beta\vert^2+
2\vert\gamma\vert^2 \right)
\right.\nonumber\\
 \fl &\quad +\left[x+\frac{1}{3}(2\beta+\gamma)\right]a_1^\dag
+\left[x+\frac{1}{3}(-\beta+\gamma)\right]a_2^\dag
\nonumber\\
 \fl &\quad +\left[x+\frac{1}{3}(-\beta-2\gamma)\right]a_3^\dag \left.
-\frac{1}{6}(a_1^\dag+a_2^\dag+a_3^\dag)^2\right\}\ket{000}
\label{tripartite}
\end{eqnarray}
This is the so called tripartite CES introduced in Ref. \cite{14}.
Using the bosonic commutative relation $[a_i,a_j^\dag]=\delta_{ij}$,
we have
\numparts
\begin{eqnarray}
 a_1 \ket{\beta,\gamma, x}_{\mu\nu\tau}
&=&\frac{1}{3\lambda}\left[\left(\beta
(\nu^2+\tau^2)+\frac{\gamma\mu\tau^2}{\nu}+3x\mu\right)\right.
\nonumber\\
& &\qquad \left. -\frac{\mu}{\lambda}\left( \mu a_1^{\dag}+\nu
a_2^{\dag}
+\tau a_3^{\dag}\right)\right]\ket{\beta,\gamma, x}_{\mu\nu\tau}\\
 a_2 \ket{\beta,\gamma, x}_{\mu\nu\tau}
&=&\frac{1}{3\lambda}\left[\left(-\beta\mu\nu+\gamma\tau^2+3x\nu\right)\right.
\nonumber\\
& &\qquad \left. -\frac{\nu}{\lambda}\left( \mu a_1^{\dag}+\nu
a_2^{\dag}
+\tau a_3^{\dag}\right)\right]\ket{\beta,\gamma, x}_{\mu\nu\tau}\\
 a_3 \ket{\beta,\gamma, x}_{\mu\nu\tau}
&=&\frac{1}{3\lambda}\left[\left(-\frac{\gamma(\mu^2+\nu^2)\tau}{\nu}-\beta\mu\tau+3x\tau\right)\right.
\nonumber\\
& &\qquad \left. -\frac{\tau}{\lambda}\left( \mu a_1^{\dag}+\nu
a_2^{\dag} +\tau a_3^{\dag}\right)\right]\ket{\beta,\gamma,
x}_{\mu\nu\tau}
\end{eqnarray}
\endnumparts
Combining the equations (10), we obtain the eigenequations of
tripartite CES $\ket{\beta,\gamma, x}_{\mu\nu\tau}$
\numparts
\begin{eqnarray}
\frac{1}{3}(\mu X_1 + \nu X_2 +\tau X_3) \ket{\beta,\gamma,
x}_{\mu\nu\tau}&=& \frac{\lambda
x}{\sqrt{2}} \ket{\beta,\gamma, x}_{\mu\nu\tau}\\
(\nu a_1 - \mu a_2)\ket{\beta,\gamma, x}_{\mu\nu\tau} &=&
\nu \beta \lambda \ket{\beta,\gamma, x}_{\mu\nu\tau}\\
(\tau a_2 - \nu a_3)\ket{\beta,\gamma, x}_{\mu\nu\tau} &=&
\tau\gamma \lambda \ket{\beta,\gamma, x}_{\mu\nu\tau}
\end{eqnarray}
\endnumparts
\par
So we see that $\ket{\beta,\gamma, x}_{\mu\nu\tau}$ is actually the
common eigenvector of $\frac{1}{3}(\mu X_1 + \nu X_2 +\tau X_3)$,
$(\nu a_1 - \mu a_2)$ and $(\tau a_2 - \nu a_3)$, and $[(\mu X_1 +
\nu X_2 +\tau X_3),(\nu a_1 - \mu a_2)]=[(\mu X_1 + \nu X_2 +\tau
X_3),(\tau a_2 - \nu a_3)]=[(\nu a_1 - \mu a_2),(\tau a_2 - \nu
a_3)]=0$.

\section{Main Properties Of $\ket{\beta,\gamma, x}_{\mu\nu\tau}$}
\label{properties}
\par
In Sec. \ref{triparite}, we construct the tripartite CES
$\ket{\beta,\gamma, x}_{\mu\nu\tau}$ just oppositely to traditional
ways, and now we will check its orthogonality and completeness, to
prove it span the Hilbert space of tripartite states, and so make up
a new kind of representation.

\par
\subsection{Orthogonal Property}

\par
We investigate whether $\ket{\beta,\gamma, x}_{\mu\nu\tau}$ is
mutual orthogonal or not. Explicitly, using the eigenequations of
tripartite CES, we examine the following matrix elements:
\numparts
\begin{eqnarray}
\fl \phantom{}_{\mu\nu\tau}\bra{\beta',\gamma', x'}\frac{\mu X_1 +
\nu X_2 + \tau X_3}{3}\ket{\beta,\gamma, x}_{\mu\nu\tau}&=&
\frac{\lambda x'}{\sqrt{2}}\phantom{n}_{\mu\nu\tau}
\braket{\beta',\gamma', x'}{\beta,\gamma, x}_{\mu\nu\tau}
\\
&=& \frac{\lambda x}{\sqrt{2}} \phantom{n}_{\mu\nu\tau}
\braket{\beta',\gamma', x'} {\beta,\gamma, x}_{\mu\nu\tau}
\end{eqnarray}
\endnumparts
which leads to
\begin{equation}
\phantom{}_{\mu\nu\tau}\braket{\beta',\gamma', x'}{\beta,\gamma,
x}_{\mu\nu\tau}(x'-x)=0
\end{equation}
\par
To derive the exact express of
$\phantom{}_{\mu\nu\tau}\braket{\beta',\gamma', x'}{\beta,\gamma,
x}_{\mu\nu\tau}$, we will use the over-completeness relation of the
three-mode coherent state
\begin{equation}
\int\frac{\rmd ^2 z_1 \rmd ^2 z_2 \rmd ^2
z_3}{\pi^3}\ketbra{z_1,z_2,z_3}{z_1,z_2,z_3}=1
\end{equation}
where
\begin{eqnarray}
\fl \ket{z_1,z_2,z_3} &= D_1(z_1)D_2(z_2)D_3(z_3)\ket{000}\nonumber\\
\fl &= \exp\left[-\frac{1}{2}\left( \vert z_1 \vert^2 +\vert z_2
\vert^2 +\vert z_3 \vert^2 \right) + z_1 a_1^\dag + z_2 a_2^\dag +
z_3 a_3^\dag\right]\ket{000}
\end{eqnarray}
and $D_i(z)=\exp(z a_i^{\dag} -z^* a_i)$. Using the definition of
Tripartite CES in \Eref{tripartite}, the overlap is
\begin{eqnarray}
\fl
& \braket{z_1,z_2,z_3}{\beta,\gamma, x}_{\mu\nu\tau}\nonumber\\
\fl &=\exp \left\{ \left[-\frac{3}{4}x^2 -\frac{1}{6\nu}
(\beta^*\gamma+ \beta\gamma^*)\mu\tau^2
-\frac{1}{6}\vert\gamma\vert^2\tau^2
\left(1+\frac{\mu^2}{\nu^2}\right) -
\frac{1}{6}\vert\beta\vert^2 \left(\nu^2+\tau^2\right)\right] \right.\nonumber\\
\fl &\quad + \frac{1}{3\lambda} \left[\left(\beta
(\mu^2+\tau^2)+\frac{\gamma\mu\tau^2}{\nu}+3x\mu\right)z_1^{*} +
\left(-\beta\mu\nu+\gamma\tau^2+3x\nu\right)z_2^{*}\right.\nonumber\\
\fl &\quad +
\left.\left(-\frac{\gamma(\mu^2+\nu^2)\tau}{\nu}-\beta\mu\tau+3x\tau\right)z_3^{*}\right]\nonumber\\
\fl &\quad  \left.-\frac{1}{6\lambda^2}\left( \mu z_1^{*}+\nu
z_2^{*}+\tau z_3^{*}\right)^2-\frac{1}{2}\left(\vert z_1 \vert^2 +
\vert z_2 \vert^2 + \vert z_3 \vert^2\right) \right\}
\end{eqnarray}
To calculate $\phantom{}_{\mu\nu\tau}\braket{\beta',\gamma',
x'}{\beta,\gamma, x}_{\mu\nu\tau}$
\begin{eqnarray}
\fl \braket{\beta',\gamma', x'}{\beta,\gamma, x} &=&\int \frac{\rmd
^2 z_1 \rmd ^2 z_2 \rmd ^2 z_3}{\pi^3}\braket{\beta',\gamma',
x'}{z_1,z_2,z_3}\braket{z_1,z_2,z_3}{\beta,\gamma,
x}\nonumber\\
&=&\exp\left\{
-\frac{\mu^2+\nu^2}{6\nu^2}\left[\nu^2(\vert\beta\vert^2+\vert\beta'\vert^2)+
\tau^2\left(\vert\gamma\vert^2+\vert\gamma'\vert^2\right)\right]\right.\nonumber\\
&&\qquad -\frac{\mu}{6\nu}\tau^2
\left[\beta\gamma^*+\beta^*\gamma+\beta'\gamma'^*+\beta'^*\gamma'-2(\beta\gamma'^*+\beta'^*\gamma)\right]\nonumber\\
&&\qquad
+\left.\frac{\nu^2+\tau^2}{3\nu^2}(\nu^2\beta\beta'^*+\mu^2\gamma\gamma'^*)
\right\}\delta(x-x') \label{orthognal}
\end{eqnarray}
In deriving \Eref{orthognal}, we have used the mathematical formula
\begin{equation}
\fl \int \frac{\rmd ^2 z}{\pi}\exp \left(\zeta\vert z\vert^2 + \xi z
+ \eta z^* + f z^2 +g z^{*2} \right) =
\frac{1}{\sqrt{\zeta^2-4fg}}\exp\left[\frac{-\zeta\xi\eta+\xi^2 g
+\eta^2 f}{\zeta^2-4fg}\right]
\end{equation}
with its convergent condition
\begin{equation*}
\mathrm{Re}(\xi+f+g) < 0,\quad
\mathrm{Re}\left(\frac{\zeta^2-4fg}{\xi+f+g}\right) < 0
\end{equation*}
or
\begin{equation*}
\mathrm{Re}(\xi-f-g) < 0,\quad
\mathrm{Re}\left(\frac{\zeta^2-4fg}{\xi-f-g}\right) < 0
\end{equation*}
and the limiting form of Dirac's delta function
\begin{equation}
\delta(x)=\lim_{\varepsilon\to{}0}\frac{1}
{\sqrt{\pi\varepsilon}}\exp\left(-\frac{x^2}{\varepsilon}\right)
\end{equation}
\par
\subsection{Completeness Relation}
\par
Now we shall check whether $\ket{\beta,\gamma,x}_{\mu\nu\tau}$
possesses the completeness relation. By virtue of the technique of
IWOP, and the normal ordered product of the three-mode vacuum
projector
\begin{equation}
\ketbra{000}{000}=:\exp(a_1^\dag a_1 + a_2^\dag a_2 + a_3^\dag a_3):
\end{equation}
we can smoothly prove the completeness relation of
$\ket{\beta,\gamma,x}_{\mu\nu\tau}$
\begin{eqnarray}
\fl &&\int\frac{\rmd ^2 \beta}{\pi} \frac{\rmd ^2 \gamma}{\pi}
\int_{-\infty}^{+\infty}\frac{\rmd  x}{\sqrt{6\pi}}
\ket{\beta,\gamma,
x}_{\mu\nu\tau\,\,\mu\nu\tau}\bra{\beta,\gamma,x}\nonumber\\
\fl &&=\int\frac{\rmd ^2 \beta}{\pi} \frac{\rmd ^2 \gamma}{\pi}
\int_{-\infty}^{+\infty}\frac{\rmd x}{\sqrt{6\pi}}:\exp
\left\{-\frac{1}{3} \left(\frac{3}{\sqrt{2}}x- \frac{\mu X_1 +\nu
X_2 +\tau X_3}{\lambda}\right)^2\right.\nonumber\\
\fl &&\qquad -\frac{1}{3} \left[\left(\nu\beta^*-\frac{\nu a_1^\dag
-\mu a_2^\dag}{\lambda}\right)\left(\nu\beta-\frac{\nu a_1 -\mu
a_2}{\lambda}\right)\right]\nonumber\\
\fl &&\qquad -\frac{1}{3} \left[\left(\tau\gamma^*-\frac{\tau
a_2^\dag -\nu a_3^\dag}{\lambda}\right)\left(\tau\gamma-\frac{\tau
a_2 -\nu
a_3}{\lambda}\right)\right]\nonumber\\
\fl &&\qquad
\left.-\frac{1}{3}\left[\left(\frac{\tau}{\nu}(\nu\beta^*
+\mu\gamma^*) -\frac{\tau a_1^\dag -\mu a_3^\dag}{\lambda}\right)
\left(\frac{\tau}{\nu}(\nu\beta +\mu\gamma) -\frac{\tau a_1 -\mu
a_3}{\lambda}\right)\right]\right\}:\nonumber\\
\fl &&=\frac{3}{\tau^2 \lambda^2}
\int_{-\infty}^{+\infty}\frac{\rmd x}{\sqrt{6\pi}}:\exp
\left[-\frac{1}{3} \left(\frac{3}{\sqrt{2}}x- \frac{\mu X_1 +\nu
X_2 +\tau X_3}{\lambda}\right)^2\right]:\\
\fl &&=\frac{1}{\tau^2 \lambda^2}
\end{eqnarray}
and also have
\begin{eqnarray}
\int\frac{\rmd ^2 \beta}{\pi}\frac{\rmd ^2 \gamma}{\pi}
\ket{\beta,\gamma, x}_{\mu\nu\tau\,\,\mu\nu\tau}\bra{\beta,\gamma,x}=\nonumber\\
\quad\frac{3}{\tau^2 \lambda^2}:\exp \left[-3
\left(\frac{1}{\sqrt{2}}x- \frac{\mu X_1 +\nu X_2 +\tau
X_3}{3\lambda}\right)^2\right]:
\end{eqnarray}
\par
\subsection{The Conjugate State of
$\ket{\beta,\gamma,x}_{\mu\nu\tau}$}
\par
According to communication  relationship between mechanic operator
and quantum state, once we known tripartite CES, we can derive its
conjugate state. Three-particle's total momentum is
$P=\Sigma_{i=1}^3P_i$, $(P_i=(a_i-a_i^\dag)/(\rmi \sqrt{2}))$, P
$(\nu a_1 -\mu a_2)$, and $(\tau a_2 -\nu a_3)$ are permutable with
each other as well, we make great efforts to find their common
eigenvector with eigenvalues $\lambda p/\sqrt{2}$,
$\nu\sigma\lambda$ and $\tau\kappa\lambda$, expressed as
$\ket{\sigma,\kappa, p}_{\mu\nu\tau}$:
\begin{eqnarray}
\fl \ket{\sigma,\kappa, p}_{\mu\nu\tau}=\exp & \left\{
\left[-\frac{3}{4}p^2 -\frac{1}{6\nu} (\sigma^*\kappa+
\sigma\kappa^*)\mu\tau^2 -\frac{1}{6}\vert\kappa\vert^2\tau^2
\left(1+\frac{\mu^2}{\nu^2}\right) -
\frac{1}{6}\vert\sigma\vert^2 \left(\nu^2+\tau^2\right)\right] \right.\nonumber\\
&\quad + \frac{1}{3\lambda} \left[\left(\sigma
(\nu^2+\tau^2)+\frac{\kappa\mu\tau^2}{\nu}+3\rmi p\mu\right)a_1^\dag
+\left(-\sigma\mu\nu+\kappa\tau^2+3\rmi p\nu\right)a_2^\dag\right.\nonumber\\
&\quad + \left.\left(-\frac{\kappa(\mu^2+\nu^2)\tau}{\nu}-\sigma\mu\tau+3\rmi p\tau\right)a_3^\dag\right]\nonumber\\
&\quad  \left.+\frac{1}{6\lambda^2}\left( \mu a_1^{\dag}+\nu
a_2^{\dag} +\tau a_3^{\dag}\right)^2 \right\}\ket{000}
\label{tri-define-conjugate}
\end{eqnarray}
The results after
annihilation operators acting on $\ket{\sigma,\kappa,
p}_{\mu\nu\tau}$ respectively are \numparts
\begin{eqnarray}
 a_1 \ket{\sigma,\kappa, p}_{\mu\nu\tau}
&=&\frac{1}{3\lambda}\left[\left(\sigma
(\nu^2+\tau^2)+\frac{\kappa\mu\tau^2}{\nu}+3\rmi p\mu\right)\right.
\nonumber\\
& &\qquad \left. +\frac{\mu}{\lambda}\left( \mu a_1^{\dag}+\nu
a_2^{\dag}
+\tau a_3^{\dag}\right)\right]\ket{\sigma,\kappa, p}_{\mu\nu\tau}\\
a_2 \ket{\sigma,\kappa, p}_{\mu\nu\tau}
&=&\frac{1}{3\lambda}\left[\left(-\sigma\mu\nu+\kappa\tau^2+3\rmi
p\nu\right)\right.
\nonumber\\
& &\qquad \left. +\frac{\nu}{\lambda}\left( \mu a_1^{\dag}+\nu
a_2^{\dag}
+\tau a_3^{\dag}\right)\right]\ket{\sigma,\kappa, x}_{\mu\nu\tau}\\
a_3 \ket{\sigma,\kappa, p}_{\mu\nu\tau}
&=&\frac{1}{3\lambda}\left[\left(-\frac{\kappa(\mu^2+\nu^2)\tau}{\nu}-\sigma\mu\tau+3\rmi
p\tau\right)\right.
\nonumber\\
& &\qquad \left. +\frac{\tau}{\lambda}\left( \mu a_1^{\dag}+\nu
a_2^{\dag} +\tau a_3^{\dag}\right)\right]\ket{\sigma,\kappa,
p}_{\mu\nu\tau}\label{conjuagatea1a2a3}
\end{eqnarray}
\endnumparts
from the above equations, we get similar expressions of its
eigenequations as those of tripartite CES
\numparts
\begin{eqnarray}
\frac{1}{3}(\mu P_1 + \nu P_2 +\tau P_3) \ket{\sigma,\kappa,
p}_{\mu\nu\tau}&=& \frac{\lambda
p}{\sqrt{2}} \ket{\beta,\gamma, x}_{\mu\nu\tau}\\
(\nu a_1 - \mu a_2)\ket{\sigma,\kappa, p}_{\mu\nu\tau} &=&
\nu \sigma \lambda \ket{\sigma,\kappa, p}_{\mu\nu\tau}\\
(\tau a_2 - \nu a_3)\ket{\sigma,\kappa, p}_{\mu\nu\tau} &=&
\tau\kappa \lambda \ket{\sigma,\kappa,
p}_{\mu\nu\tau}\label{tri-eigen-conjugate}
\end{eqnarray}
\endnumparts
So far we get tripartite momentum CES. Using the IWOP technique we
can prove
\begin{eqnarray}
\int\frac{\rmd ^2 \sigma}{\pi} \frac{\rmd ^2
\kappa}{\pi}\ket{\sigma,\kappa, p}_{\mu\nu\tau\,\,\mu\nu\tau}
\bra{\sigma,\kappa, p}=\nonumber\\
\quad\frac{3}{\tau^2
\lambda^2}:\exp\left[-3\left(\frac{1}{\sqrt{2}}p-\frac{ \mu P_1+\nu
P_2 + \tau P_3}{3\lambda}\right)^2\right]:
\end{eqnarray}
so the completeness integration also holds, i.e.,
\begin{equation}
\int_{-\infty}^{\infty}\frac{\rmd p}{\sqrt{6\pi}}\int\frac{\rmd ^2
\sigma}{\pi} \frac{\rmd ^2 \kappa}{\pi}\ket{\sigma,\kappa,
p}_{\mu\nu\tau\,\,\mu\nu\tau} \bra{\sigma,\kappa, p}=
\frac{1}{\tau^2 \lambda^2}
\end{equation}

Thus $\ket{\sigma,\kappa, p}_{\mu\nu\tau\,\,\mu\nu\tau}$ is the
conjugate state of $\ket{\beta,\gamma, x}_{\mu\nu\tau}$.
\section{Generating Tripartite CES By Asymmetric Beam-splitter(BS)}
\label{generating-CES}
\par
The tripartite CES as we show upper, can be generated by asymmetric
BS operator. One function of BS it to generate entangled state
\cite{19}, and operator representation of BS operating on incident
optic field can be expressed (with phase-free) by\cite{20}
\begin{equation}
B_{ij}(\theta)=\exp\left[-\theta\left(a_i^\dag a_j -a_i
a_j^\dag\right)\right]
\end{equation}
\par
Letting the ideal single-mode maximal-squeezed state in mode 1,
expressed by
$\ket{x=0}_1=exp\left[-\frac{1}{2}a_1^\dag\right]\ket{0}_1$, and the
vacuum state $\ket{0}_{2,3}$ in mode $2$, $3$ respectively enter the
two input ports of two sequential asymmetric BS and get overlapped,
we have
\begin{eqnarray}
\fl
B_{23}(\varphi)B_{12}(\theta)&\exp\left[-\frac{1}{2}a_1^\dag\right]
\ket{0}_1\otimes\ket{0}_{2}\otimes\ket{0}_{3}=\nonumber\\
\fl &\exp\left[-\frac{1}{2}\left(a_1^\dag \cos(\theta)+a_2^\dag
\sin(\theta) \cos(\varphi)+a_3^\dag
\sin(\theta)\sin(\varphi)\right)^2\right]\ket{000}
\label{asymmetry-bs} \end{eqnarray}

\par
Since
$\cos^2(\theta)+\sin^2(\theta)\cos^2(\varphi)+\sin^2(\theta)\sin^2(\varphi)=1$,
so $\lambda$ is introduced as $3\lambda^2=\mu^2+\nu^2+\tau^2$ as
illustrated in Sec. \ref{triparite}. When
$\theta=\arccos(\frac{\mu}{\sqrt{3}\lambda})$ and
$\varphi=\arccos(\frac{\nu}{\sqrt{\nu^2+\tau^2}})$, the state out of
the two sequential BS in \Eref{asymmetry-bs} becomes
\begin{equation}
\exp\left[-\frac{1}{6\lambda^2}\left(\mu a_1^\dag +\nu a_2^\dag +
\tau a_3^\dag \right)^2\right]\ket{000}_{123}
\end{equation}
which is a three-mode squeezed state. Then operating three
sequential displacement operators $D_1(\epsilon_1)$,
$D_2(\epsilon_2)$, $D_3(\epsilon_3)$ on three individual mode, where
$D_i(\epsilon)$ writes
\begin{equation}
D_i(\epsilon)=\exp(\epsilon a_i^\dag -\epsilon^* a_i)
\end{equation}
and the displacements $\epsilon_1$, $\epsilon_2$, $\epsilon_3$ are
\numparts
\begin{eqnarray}
\epsilon_1&=&\frac{\phantom{-}2\beta \nu(\nu^2+\tau^2)+2\gamma\mu\tau^2+3x\mu\nu}{6\nu\lambda}\\
\epsilon_2&=&\frac{-2\beta\mu\nu + 2\gamma\tau^2 +3x\nu}{6\lambda}\\
\epsilon_3&=&\frac{-2\beta\mu\nu\tau-2\gamma\tau(\mu^2+\nu^2)+3x\nu\tau}{6\nu\lambda}
\label{epsilon3}
\end{eqnarray}
\endnumparts
\par
After these three sequential displacements, the ideal three-mode
asymmetry squeezed state will becomes
\begin{eqnarray}
\fl D_1(\epsilon_1)D_2(\epsilon_2)D_3(\epsilon_3)&
\exp\left[-\frac{1}{6\lambda^2}\left(\mu a_1^\dag +\nu a_2^\dag +
\tau a_3^\dag \right)^2\right]\ket{000}\nonumber\\
=&\exp\left\{-\frac{\epsilon_1\epsilon_1^*+\epsilon_2\epsilon_2^*+
\epsilon_3\epsilon_3^*}{2}+
\epsilon_1 a_1^\dag+\epsilon_2 a_2^\dag+\epsilon_3 a_3^\dag\right.\nonumber\\
& \qquad \left. -\frac{1}{6\lambda^2}\left(\mu (a_1^\dag-
\epsilon_1^*) +\nu (a_2^\dag- \epsilon_2^*) + \tau (a_3^\dag-
\epsilon_3^*) \right)^2\right\}\ket{000}\label{generate}
\end{eqnarray}

Substitute the expression of $\epsilon_i$ into \Eref{generate}, we
find the state is tripartite CES compared with \Eref{tripartite}.
And experimentally, we can achieve these displacements (eg.
$D_1(\epsilon_1)$), by reflecting the light field
 $\exp\left[-\frac{1}{6\lambda^2}\left(\mu a_1^\dag +\nu a_2^\dag +
\tau a_3^\dag \right)^2\right]\ket{000}$ from a partially reflecting
mirror (say $99\%$ reflection and $1\%$ transmission) and adding
through the mirror a field that has been phase and amplitude
modulated according to the values $\mu$, $\nu$, $\tau$, and $\beta$,
$\gamma$, $x$. Thus the tripartite CES $\ket{\beta,\gamma,
x}_{\mu\nu\tau}$ can be implemented.

\section{Some Applications of $\ket{\beta,\gamma, x}_{\mu\nu\tau}$}
\label{application}
\par
In this section, we briefly introduce some probably applications of
the tripartite CES.
\par
\subsection{Wigner Operator}
\par
In atomic and quantum optic area, Wigner distribution as the
quasi-classical distribution \cite{21,22} well represent the
non-classical properties of quantum state through its partial
negativity in quadrature phase. One basic way to obtain Wigner
distribution is to trace production between matrix density and
Wigner operator \cite{23}. Analogue to single mode Wigner function,
and basing on the completeness and orthogonality of tripartite CES,
we now introduce the following ket-bra integration
\begin{equation}
\fl \int\frac{\rmd ^2 \beta}{\pi} \frac{\rmd ^2 \gamma}{\pi}
\int_{-\infty}^{+\infty}\frac{\rmd  u}{2\pi\sqrt{6\pi}}e^{3\rmi p
u/2} \ket{\beta,\gamma,
x+\frac{u}{2}}_{\mu\nu\tau\,\,\mu\nu\tau}\bra{\beta,\gamma,x-\frac{u}{2}}\equiv
\triangle(p,x)
\end{equation}
\par
Considering the explicit definition of $\ket{\beta,\gamma,
x}_{\mu\nu\tau}$ in \Eref{tripartite} and employ the IWOP technique,
we can directly calculate out
\begin{eqnarray}
\fl \triangle(p,x)=\frac{1}{\pi\tau^2 \lambda^2} :\exp
&\left[-3\left(\frac{x}{\sqrt{2}}-\frac{ \mu X_1+\nu X_2 + \tau
X_3}{3\lambda}\right)^2 \right.\nonumber\\
&\quad\left.-3\left(\frac{p}{\sqrt{2}}-\frac{ \mu P_1+\nu P_2 + \tau
P_3}{3\lambda}\right)^2\right]:
\end{eqnarray}
\par
which is a generalization of the normally ordered form of the usual
Wigner operator.  We may integrate $\Delta(p,x)$ out of \emph{x}
 \emph{p}, respectively, e.g.
 \numparts
\begin{eqnarray}
\fl \int_{-\infty}^{+\infty}\rmd  x
\Delta(p,x)&=&\sqrt{\frac{2}{3\pi}}\frac{1}{\tau^2
\lambda^2}:\exp\left[-3\left(\frac{1}{\sqrt{2}}p-\frac{ \mu P_1+\nu
P_2 + \tau P_3}{3\lambda}\right)^2\right]:\\
\fl &=& \frac{1}{9}\sqrt\frac{6}{\pi} \int\frac{\rmd ^2 \sigma}{\pi}
\frac{\rmd ^2 \kappa}{\pi}\ket{\sigma,\kappa,
p}_{\mu\nu\tau\,\,\mu\nu\tau}
\bra{\sigma,\kappa, p} \label{wigner-p}\\
\fl \int_{-\infty}^{+\infty}\rmd  p \Delta(p,x)&= &
\sqrt{\frac{2}{3\pi}}\frac{1}{\tau^2
\lambda^2}:\exp\left[-3\left(\frac{1}{\sqrt{2}}x-\frac{ \mu X_1+\nu
X_2 + \tau X_3}{3\lambda}\right)^2\right]:\\
\fl &= &  \frac{1}{9}\sqrt\frac{6}{\pi} \int\frac{\rmd ^2
\beta}{\pi} \frac{\rmd ^2 \gamma}{\pi}\ket{\beta,\gamma,
x}_{\mu\nu\tau\,\,\mu\nu\tau} \bra{\beta,\gamma,x} \label{wigner-x}
\end{eqnarray}
\endnumparts
\par
Following Wigner's original idea of setting up a function in $x$-$p$
phase whose marginal distribution is the probability of finding a
particle in coordinate space and momentum space, respectively. we
can immediately judge that the wigner operator $\Delta(p,x)$ in
equations \eref{wigner-x} and \eref{wigner-p} is just a marginal
distributional Wigner operator. Then the marginal distribution in
the $p$-direction and its conjugate marginal distributions in the
$x$-direction are \numparts
\begin{eqnarray}
&&\int_{-\infty}^{+\infty}\rmd  x
\bra{\psi}\Delta(p,x)\ket{\psi}\nonumber\\
&=&\sqrt{\frac{2}{3\pi}}\frac{1}{\tau^2
\lambda^2}\bra{\psi}:\exp\left[-3\left(\frac{1}{\sqrt{2}}p-\frac{
\mu P_1+\nu
P_2 + \tau P_3}{3\lambda}\right)^2\right]:\ket{\psi}\nonumber\\
&=& \sqrt\frac{1}{6\pi} \int\frac{\rmd ^2 \sigma}{\pi} \frac{\rmd ^2
\kappa}{\pi}\left\vert\braket{\psi}{\sigma,\kappa, p}_{\mu\nu\tau} \right\vert^2 \\
&&\int_{-\infty}^{+\infty}\rmd  p
\bra{\psi}\Delta(p,x)\ket{\psi}\nonumber\\
&= &
\sqrt{\frac{2}{3\pi}}\frac{1}{\tau^2 \lambda^2}
\bra{\psi}:\exp\left[-3\left(\frac{1}{\sqrt{2}}x-\frac{ \mu X_1 +
\nu X_2 + \tau X_3}{3\lambda}\right)^2\right]:\ket{\psi}\nonumber\\
&= &  \sqrt\frac{1}{6\pi} \int\frac{\rmd ^2 \beta}{\pi} \frac{\rmd
^2 \gamma}{\pi}\left\vert\braket{\psi}{\beta,\gamma, x}_{\mu\nu\tau}
\right\vert^2
\end{eqnarray}
\endnumparts
correspondingly. Furthermore, we should note that in this case the
classical $x$-$p$ phase space corresponds to the operators $X$ and
$P$ respectively.

\par
\subsection{Three-Mode Squeezing Operator}
\label{squeeze}
\par
One important application of IWOP technique is to construct squeezed
operator no matter how complex the quantum states is in continuous
variable quadrature space \cite{26,27}. In a similar way, We take a
classical transformation $x \to \frac{x}{\eta}$ in
$\ket{\beta,\gamma, x}_{\mu\nu\tau}$ to build a ket-bra integral,
\begin{equation}
S(\eta)=\tau^2\lambda^2\int\frac{1}{\pi^2}\rmd ^2 \beta \rmd ^2
\gamma \int_{-\infty}^{+\infty}\frac{\rmd  x}{\sqrt{6\eta\pi}}
\ket{\beta,\gamma,
x/\eta}_{\mu\nu\tau\,\,\mu\nu\tau}\bra{\beta,\gamma,x}
\label{squeezed}
\end{equation}
\par
Using the IWOP technique, we can directly perform the integration in
\Eref{squeezed} to obtain
\begin{eqnarray}
\fl S(\eta)&=&\sech^{1/2}(\zeta)\exp \left\{-\frac{1}{6\lambda^2}
(\mu a_1^\dag + \nu a_2^\dag + \tau a_3^\dag)^2
\tanh\zeta\right\} \nonumber \\
\fl &&:\exp\left\{\frac{1}{3\lambda^2}(\sech\zeta -1) (\mu a_1^\dag
+\nu
a_2^\dag + \tau a_3^\dag ) (\mu a_1+ \nu a_2+\tau a_3)\right\}:\nonumber\\
\fl &&\exp\left\{\frac{1}{6\lambda^2}(\mu a_1+ \nu a_2+\tau a_3)^2
\tanh\zeta\right\} \label{squeezing-operator}
\end{eqnarray}
which is a new three-mode squeezing operator with parameter $\eta$,
and where $\eta=\exp(\zeta)$, $\sech\zeta=2\eta/(1+\eta^2)$ and
$\tanh\zeta=(\eta^2-1)/(1+\eta^2)$. To make this squeezing more
compact, we introduce the notation $R^\dag=\frac{\mu a_1^\dag + \nu
a_2^\dag + \tau a_3^\dag}{{\sqrt{3}}\lambda}$, and using the
following formula $:\exp[(e^\zeta -1)a^\dag a]:=\exp(\zeta a^\dag
a)$, we can rewrite the \Eref{squeezing-operator} as
\begin{eqnarray}
S(\eta)&=&\sech^{1/2}(\zeta)\exp \left\{-\frac{1}{2}R^{\dag 2}
\tanh\zeta\right\}\nonumber\\
&&\times:\exp\left\{ (\sech\zeta -1)R^\dag R \right\}:
\exp\left\{\frac{1}{2} R^2
\tanh\zeta\right\} \nonumber\\
 &=&\sech^{1/2}(\zeta)\exp
\left\{-\frac{1}{2}R^{\dag 2}
\tanh\zeta\right\}\nonumber\\
&&\times\exp\left\{ R^\dag R \ln\sech \zeta \right\}
\exp\left\{\frac{1}{2} R^2 \tanh\zeta\right\}
\label{squeezing-operator-compact}
\end{eqnarray}
\par
And we can also find that $R$ compose a $SU(1,1)$ Lie algebra as
\begin{equation}
[ R, R^\dag]=1,\qquad [\frac{1}{2}R^2,\frac{1}{2}R^{\dag
2}]=R^{\dag}R+\frac{1}{2}
\end{equation}
The squeezing operator $S(\eta)$ squeezes states $\ket{\beta,\gamma,
x}_{\mu\nu\tau}$ in a natural way
\begin{equation}
S(\eta)\ket{\beta,\gamma, x}_{\mu\nu\tau} =
\frac{1}{\sqrt{\eta}}\ket{\beta,\gamma, x/\eta}_{\mu\nu\tau}
\end{equation}
 Correspondingly, the three-mode squeezed vacuum state is
\begin{equation}
\fl S(\eta)\ket{000}=\sech^{1/2}(\lambda)\exp
\left\{-\frac{1}{6}(\mu a_1^\dag + \nu a_2^\dag + \tau a_3^\dag)^2
\tanh\lambda\right\}\ket{000}
\end{equation}

Using \Eref{squeezing-operator-compact} and the Baker-Hausdroff
formula
\begin{equation}
e^ABe^{-A}=B+[A,B]+\frac{1}{2!}[A,[A,B]]+\frac{1}{3!}[A,[A,[A,B]]]+\cdots
\end{equation}
we see that
\begin{equation}
S(\eta)a_iS(\eta)^{-1}=a_i
+\frac{\mu_i}{\sqrt{3}\lambda}\left[R(\cosh\zeta - 1)+ R^\dag
\sinh\zeta\right]
\end{equation}
where $\mu_{1,2,3}=\mu,\nu,\tau$. It then follows from \Eref{xp}
that
\numparts
\begin{eqnarray}
S(\eta)X_iS(\eta)^{-1}=X_i+\mu_i A(e^\zeta -1)\\
S(\eta)P_iS(\eta)^{-1}=P_i+\mu_i B(e^{-\zeta} -1)
\end{eqnarray}
\endnumparts
and $A=\sum_j\mu_j X_j/(3\lambda^2)$, $B=\sum_j\mu_j
P_j/(3\lambda^2)$. So, under the $S(\eta)$ transformation the three
quadrtures of the three-mode optical field become \numparts
\begin{eqnarray}
\fl S(\eta)(X_1+X_2+X_3)S(\eta)^{-1}&=
X_1+X_2+X_3+(\mu+\nu+\tau) A(e^\zeta -1)\\
\fl S(\eta)(P_1+P_2+P_3)S(\eta)^{-1}&= P_1+P_2+P_3+(\mu+\nu+\tau)
B(e^{-\zeta} -1)
\end{eqnarray}
\endnumparts

Operating $S(\eta)^{-1}$ on the three-mode vacuum state, we obtain
the squeezed vacuum state
\begin{equation}
S(\eta)^{-1}\ket{000}=\sech^{1/2}\zeta\exp
\left[\frac{\tanh\zeta}{2}R^{\dag 2}\right]\ket{000}\equiv
\ket{\ }_{ \rho}
\end{equation}

The expectation values of the two quadratures in this state are
\begin{equation}
{}_\rho\bra{\ }(X_1+X_2+X_3)\ket{\ }_\rho=0,\qquad {}_\rho
\bra{\ }(P_1+P_2+P_3)\ket{\ }_\rho=0
\end{equation}
thus the variance of the two quadrature are
\numparts
\begin{eqnarray}
{}_\rho\bra{\ }\Delta(X_1+X_2+X_3)^2\ket{\ }_\rho&={}_\rho\bra{\
}(X_1+X_2+X_3)^2\ket{\ }_\rho\\
&=\frac{1}{2}\left[\frac{(\mu+\nu+\tau)^2}{3\lambda^2}(e^{2\zeta}-1)+3\right]\label{delta-x}\\
{}_\rho \bra{\ }\Delta(P_1+P_2+P_3)^2\ket{\ }_\rho&={}_\rho \bra{\
}(P_1+P_2+P_3)^2\ket{\ }_\rho\\
&=\frac{1}{2}\left[\frac{(\mu+\nu+\tau)^2}{3\lambda^2}(e^{-2\zeta}-1)+3\right]\label{delta-p}
\end{eqnarray}
\endnumparts
and the minimum uncertainty relation is
\begin{eqnarray}
\fl \quad \Delta(X_1+X_2+X_3)^2\Delta(P_1+P_2+P_3)^2=\nonumber\\
\fl \qquad
\frac{1}{4}\left[\frac{(\mu+\nu+\tau)^2}{3\lambda^2}(e^{2\zeta}-1)+3\right]\left[\frac{(\mu+\nu+\tau)^2}{3\lambda^2}(e^{-2\zeta}-1)+3\right]
\end{eqnarray}
In particular, when $\mu=\nu=\tau=1$, the squeezed vacuum state  $
\ket{\ }_\rho $ reduces to the usual three-mode squeezed vacuum
state, equations \eref{delta-x}, \eref{delta-p},respectively, become
\begin{eqnarray}
{}_\rho\bra{\ }\Delta(X_1+X_2+X_3)^2\ket{\ }_\rho=\frac{3}{2}e^{2\zeta}\\
{}_\rho \bra{\ }\Delta(P_1+P_2+P_3)^2\ket{\ }_\rho=\frac{3}{2}e^{-2\zeta}\\
\Delta(X_1+X_2+X_3)\Delta(P_1+P_2+P_3)=\frac{9}{4}
\end{eqnarray}
as expected. On the other hand, due to $3\lambda^2\geq
\mu\nu+\nu\tau+\mu\tau$, i.e. $(3\lambda)^2\geq(\mu+\nu+\tau)^2$.
For $\zeta\geq 0$, from equations \eref{delta-x}and \eref{delta-p},
we have \numparts
\begin{eqnarray}
{}_\rho\bra{\ }\Delta(X_1+X_2+X_3)^2\ket{\ }_\rho&=
\frac{1}{2}\left[\frac{(\mu+\nu+\tau)^2}{3\lambda^2}(e^{2\zeta}-1)+3\right]\\
&\leq\frac{3}{2}e^{2\zeta}\\
{}_\rho \bra{\ }\Delta(P_1+P_2+P_3)^2\ket{\ }_\rho&=
\frac{1}{2}\left[\frac{(\mu+\nu+\tau)^2}{3\lambda^2}(e^{-2\zeta}-1)+3\right]\\
&\geq\frac{3}{2}e^{-2\zeta}
\end{eqnarray}
\endnumparts
which implies that the squeezed vacuum state $\ket{\ }_\rho$  may
exhibit stronger squeezing in one quadrature than that of the usual
two-mode squeezed vacuum state while exhibiting weaker squeezing in
another quadrature.

\section{Conclusion}
In summary, we have brought out the ways to construct tripartite CES
just contrary the traditional method and check it correckness. We
analyzed some major properties of the tripartite CES, i.e. the
completeness relation and partly orthogonality. And a simple
experimental protocol to produce tripartite CES was also proposed by
using an asymmetric BS, which provides a new way to predict new
tripartite squeezed operator.

\ack This work supported by the President Foundation of Chinese
Academy of Science and National Natural Science Foundation of China
under grant 10475057.

\end{document}